# Tuning the circularly polarized photoluminescence of chiral 2D perovskites by high pressure

*Shenyu Dai*

**abstract**:

Chiral 2D perovskites are of great interest as circularly polarized photoluminescence materials, but these materials generally exhibit weak CPL under ambient conditions. Several studies have shown that the degree of CPL can be enhanced by using strong external magnetic fields or low temperature. Here we report a method to tune the circularly polarized photoluminescence of chiral 2D perovskites by utilizing extreme high pressure. The (S- and R-MBA)$_2$PbI$_4$ perovskites exhibited good optical tunability with pressure in term of PL wavelength, intensity and bandgap. Polarization-resolved photoluminescence measurements shown that the degree of CPL was increased from nearly zero at ambient pressure to as high as 10% at 8.5 GPa. ADXRD and Raman results indicated that the structural distortion and Increased interlayer coupling are responsible for the enhanced chirality when pressure is applied. Our findings provide a new approach to adjusting CPL materials and show potential applications in next generation of CPL devices.

1. Introduction

Polarization is one of the fundamental properties of light and represents the geometrical orientation of transverse waves. When the phases of two orthogonal polarization directions are retarded by 90°, so-called circular polarization is formed. Light with circular polarization has been widely used in 3D displays[1,2], spintronics[3,4], information storage[5], asymmetric catalysis[6] and photoelectric devices[7,8]. Conventionally the realization of circularly polarized light requires the use of various optical structures such as polarizers and wave plates, which is not conducive for integration and miniaturization of devices and will lose at least 50% of the energy. Therefore, obtaining circularly polarized photoluminescence directly from the designed materials has been attracting enormous attention[9-11] for its potential application in next generation optical and spintronic devices.

Chiral 2D perovskites are of great interest as a type of circularly polarized photoluminescence materials, which combine the advantages of 2D perovskites and chiral organic molecules[12-14]. Compared with their 3D counterparts, 2D Ruddlesden–Popper lead perovskites generally exhibit high PL quantum efficiency, strong quantum confinement effect, high absorption coefficients and better stability[15], which are beneficial for optoelectronic devices such as LED[16], photodetectors[17] and solar cells[18]. When incorporating the chiral organic molecules as interlayers, 2D perovskites will have noncentrosymmetric structures and exhibit chirality[19]. A large number of different chiral 2D perovskites have been reported exhibit interesting physical properties, such as circular dichroism (CD)[20], circularly polarized photoluminescence (CPL)[10,21,22], second-harmonic generation (SHG)[23] and bulk photovoltaic effects (BPVE)[24]. However, the CPL degree of chiral 2D perovskites is generally small at ambient condition, for example 3% for RDCPs (⟨n⟩ = 2) perovskites. The degree of CPL can be significantly enhanced by using external magnetic field[25] or reduce the temperature[10]. A record photoluminescence polarization of 17.3% was observed at 77 K. With increasing temperature, the degree of photoluminescence polarization decreases substantially due to the enhanced electron–phonon couplings and thermal-expansion interactions, which reduce the lattice distortion and, thus, decrease the chirality.

Pressure, as a unique thermodynamic variable, provides a powerful means to study the structural and electronic behaviors of lead halides perovskite materials. The optical properties of low-dimensional perovskites under high pressure have been widely investigated, and it has been found that pressure will induce great lattice distortion in perovskite materials, leading to unique phenomena such as pressure-induced broadband emission[26], broad PL tunability[27] and great PLQY enhancement[28]. It is interesting to relate the pressure induced lattice distortion with the chirality of chiral 2D perovskites.

Here we report a method utilizing pressure to tune the circularly polarized photoluminescence of chiral 2D perovskites. Chiral 2D perovskite (S- and R-MBA)$_2$PbI$_4$ crystals are systematically studied with applied pressure up to 20 GPa at room temperature by conducting in situ high-pressure PL, polarization-resolved PL, UV–vis absorption, X-ray diffraction, Raman experiments. The (S- and R-MBA)$_2$PbI$_4$ perovskites exhibit good optical pressure-tunability in term of PL wavelength, intensity and band structures. Polarization-resolved photoluminescence measurements show that the degree of CPL is increased from nearly negligible at ambient pressure to as high as 10% at 8.5 GPa. ADXRD and Raman results indicate that the structural distortion and Increased interlayer coupling are responsible for the enhanced chirality when pressure is applied.

## 2. Results and discussion

To investigate the pressure effects on the chirality of chiral perovskites, widely studied chiral 2D perovskite (S- and R-MBA)$_2$PbI$_4$ crystals (MBA = C$_6$H$_5$C$_2$H$_4$NH$_3$) with a high degree of circularly polarized PL and CPL detectability are selected (Fig.1a). In (S-MBA)$_2$PbI$_4$ and (R-MBA)$_2$PbI$_4$, chiral organic interlayers have a large mole fraction of about 67% and thus exhibit stronger chirality compared with the reduced-dimensional perovskites. Chiral 2D perovskites were firstly synthesized by incorporating MBA with different chirality into perovskite lattice, similar to previous reported method[29]. The synthesized samples with different left- and right-handed cations are both orange-colored flakes. And the phase purity of these samples was demonstrated by X-Ray diffraction results which show a series of periodic peaks.

All the in situ high pressure optical measurements were performed in a symmetric diamond anvil cell (DAC) with a culet size of 500 μm. Selected high quality (S- and R-MBA)$_2$PbI$_4$ microflakes and ruby sphere (for pressure calibration) were sealed into the 150 μm-diameter DAC chamber, using silicone oil as pressure transmission media. First, we investigated the pressure-dependent photoluminescence (PL) properties of these chiral 2D perovskites by using a 405 nm continuous wave laser. We found that (S- and R-MBA)$_2$PbI$_4$ have almost the same PL behavior when applying pressure, although they have different chirality. Here we show the PL properties of (R-MBA)$_2$PbI$_4$ in Fig.1 and (S-MBA)$_2$PbI$_4$ in SI. Despite their chirality, (S- and R-MBA)$_2$PbI$_4$ also retain the characteristics of ultrabroad energy tunability as 2D perovskites, as shown in Fig.1b and SI. At ambient pressure, (R-MBA)$_2$PbI$_4$ shows a typical 2D perovskite narrow band emission with a PL peak wavelength at 517 nm. The long tail in the PL may come from self-trapped states and defects. With the increase of pressure, the PL intensity significantly increased by about 10 times at 2 GPa, then gradually decreased until it could not be resolved beyond 9.2 GPa. Meanwhile, the peak wavelength continuously red shifted from 517 nm to 627 nm (Fig.1c), which is almost linear with the pressure increase (about 12.5 nm/GPa). This wide color tuning can be clearly seen in CIE diagram (Fig.1d), the PL coordinates change from green (1 atm), yellow (4.50 GPa), orange (6.23 GPa) to red (9.24 GPa). It worth noting that these PL coordinates are close to the edge of CIE diagram, which is beneficial to realize a wide color gamut display device.

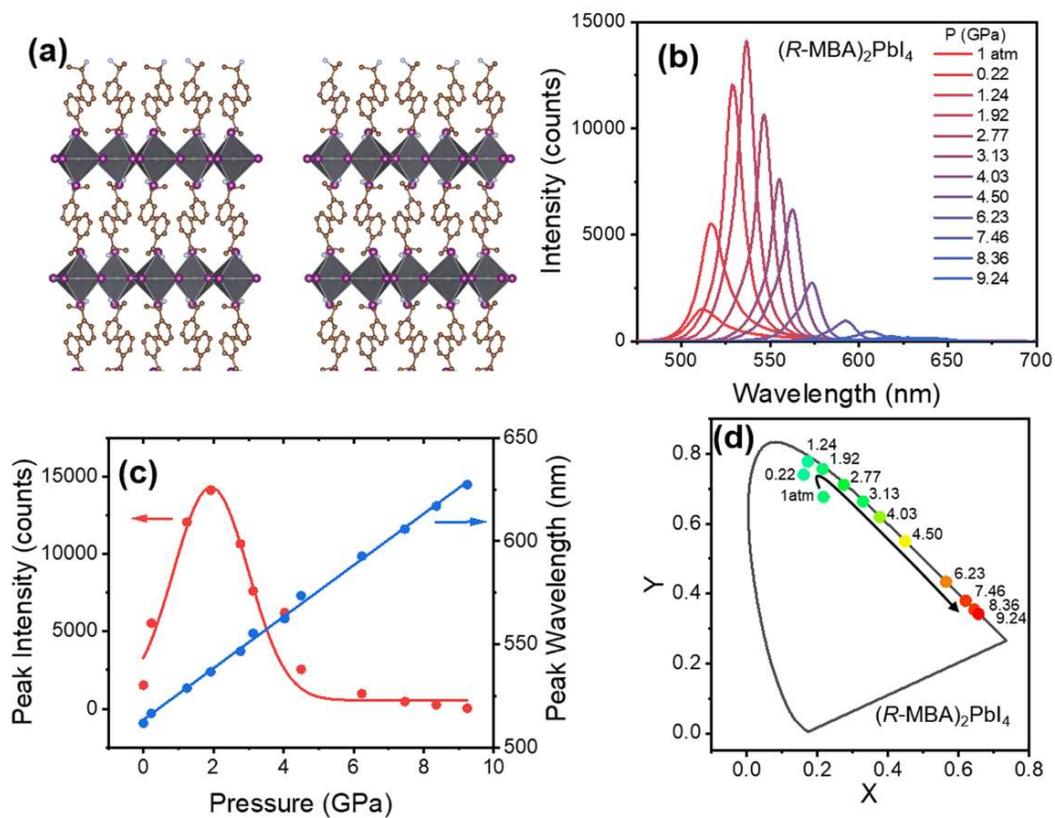

Figure 1. (a) Schematic crystal structure of (S-MBA)$_2$PbI$_4$ and (R-MBA)$_2$PbI$_4$. (b) In situ PL spectra of (R-MBA)$_2$PbI$_4$ under high pressure. (c) The peak intensity of (R-MBA)$_2$PbI$_4$ as a funciton of pressure (red) and peak wavelength as a function of pressure (blue). (d) Pressure-dependent CIE coordinates of the PL emissions of (R-MBA)$_2$PbI$_4$.

The relationship between the band structure and the pressure was also investigated. Similarly, the absorption spectra of (S-MBA)$_2$PbI$_4$ and (R-MBA)$_2$PbI$_4$ are almost identical. Here we show absorption spectra of (R-MBA)$_2$PbI$_4$ as a function of pressure in Fig.2a. At ambient pressure, an absorption peak can be observed at 500 nm. The absorption peak continuously redshifted to longer wavelength with the increasing pressure, consistent with the pressure dependent PL behavior, indicating that they are all associated with free excitons. Meanwhile, the absorption peak gradually broadened until the peak became indistinguishable when the pressure increased to 11.84 GPa, which was due to the pressure-induced lattice disorder. By Tauc plot method, the optical bandgap of (R-MBA)$_2$PbI$_4$ under different pressure was determined and shown in Fig.2c. The bandgap of (R-MBA)$_2$PbI$_4$ was approximately linearly decreased with pressure from 2.387 eV at 1 atm to 1.926 eV at 10.40 GPa, with a slope of about 44 meV/GPa. This apparent change in bandgap can be visualized intuitively as sample piezochromic from the micrograph of the DAC chamber, as shown in Fig.2b. The color of (R-MBA)$_2$PbI$_4$ changed from yellow (1 atm), orange (2.3 GPa) to red (4.8 GPa) and finally opaque at 11.40 GPa. These properties again show that this 2D chiral perovskite has good optical pressure-tunability.

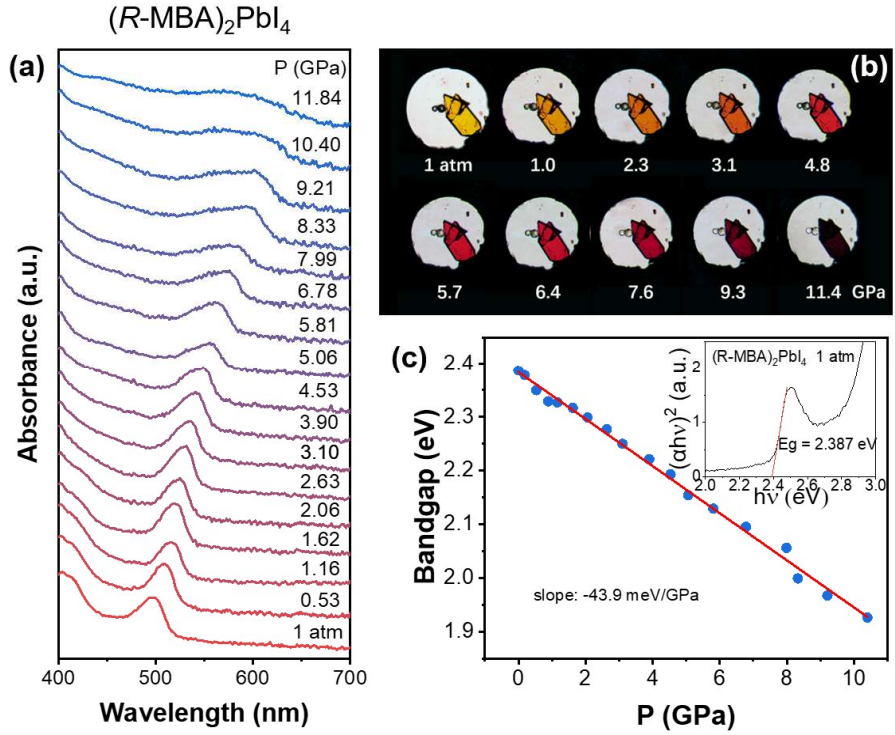

Figure 2. (a) Optical absorption spectra of (R-MBA)$_2$PbI$_4$ as a function of pressure. (b) Micrograph of (R-MBA)$_2$PbI$_4$ microflake under different pressure. The diameter of the chamber hole is 150 μm. (c) Bandgap of (R-MBA)$_2$PbI$_4$ with increasing pressure. The inset shows the tauc plot of (R-MBA)$_2$PbI$_4$ at amtomsphere pressure.

We further studied the circularly polarized photoluminescence properties of (S- and R-MBA)$_2$PbI$_4$ under high pressure. In order to obtain circularly polarized PL, a polarization-resolved confocal microscope was used, and the schematic of optical setup is shown in Fig.3a. A high stable Linearly polarized 405 nm CW laser was used as excitation. Though a quarter-wave plate and an analyzer, Left-handed (σ$^-$) and right-handed (σ$^+$) CPL from samples in the DAC chamber could be distinguished and recorded by a CCD detector. To avoid polarization effects on detector which can lead to measurement error, a half-wave plate was placed before the analyzer. The half-wave plate was mounted on a motorized rotation stage to ensure the accuracy of each measurement. We have carefully calibrated our system in advance to rule out any possible artifacts by using a depolarized halogen tungsten light.

Circularly polarized PL spectra of (S-MBA)$_2$PbI$_4$ and (R-MBA)$_2$PbI$_4$ under different pressure are shown in Fig.3b. PL intensities at each pressure are normalized by corresponding σ$^+$ intensity for comparison. At ambient pressure, no discernible difference between the σ$^-$ and σ$^+$ CPL can be found for both (S- and R-MBA)$_2$PbI$_4$. However, when pressure is increased to about 4 GPa, obvious intensity difference between σ$^-$ and σ$^+$ CPL can be observed, and this difference has opposite signs for different chiral perovskite. For (S-MBA)$_2$PbI$_4$, σ$^+$ CPL is larger than σ$^-$ CPL, but for (R-MBA)$_2$PbI$_4$, σ$^+$ CPL is weaker than σ$^-$ CPL. It is important evidence that intensity difference originates from the chirality of the material, rather than the influence of the measurement system. The relative intensity difference is larger when the pressure is further increased to about 8 GPa, although the PL intensity at this pressure becomes very weak.

To quantify the effect of pressure on circularly polarized PL spectra, the degree of polarization $P$ was calculated by following definition:

$$P = \frac{I_{\sigma^+} - I_{\sigma^-}}{I_{\sigma^+} + I_{\sigma^-}}$$

where $I_{\sigma^+}$ and $I_{\sigma^-}$ are the areas of the left- and right-handed CPL emission under photoexcitation, respectively. Fig.3c displays the relationship between the degree $P$ value and the pressure for (S-MBA)$_2$PbI$_4$ and (R-MBA)$_2$PbI$_4$. For both (S-MBA)$_2$PbI$_4$ and (R-MBA)$_2$PbI$_4$ microflakes, the absolute value of $P$ increases approximately linearly with the pressure but with opposite signs, suggesting that they have different chirality. The $P$ value is as high as 10% when pressure is increased to about 8.5 GPa. The significant increase in $P$ value suggests that pressure can be used not only to control the PL intensity and peak wavelength of a material, but also to tune the degree of the circularly polarized photoluminescence.

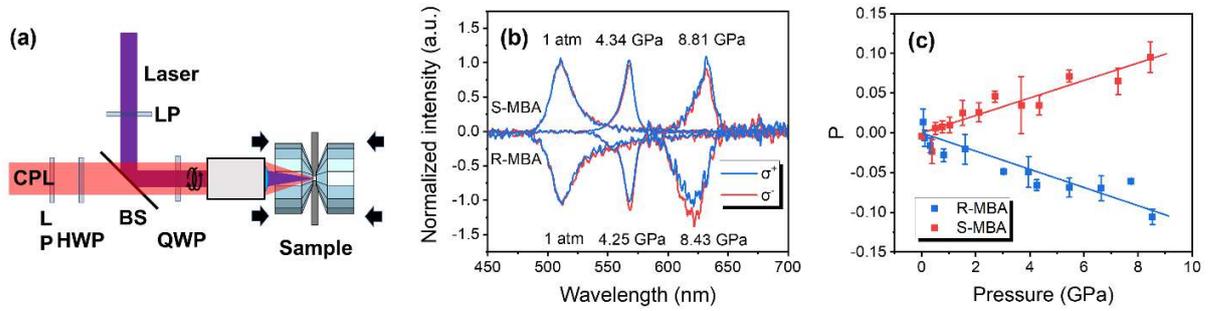

Figure 3. (a) Optical setup for polarization-resolved photoluminescence measurements. The optical components: linear polarizer (LP), quarter-wave plate (QWP), half-wave plate (HWP), beam splitter (BS). (b) Circularly polarized PL spectra of (S-MBA)$_2$PbI$_4$ and (R-MBA)$_2$PbI$_4$ excited by a 405 nm laser under different pressure. σ$^-$ (blue) and σ$^+$ (red) denoted left-handed and right-handed circularly polarized PL, respectively. PL intensities at each pressure are normalized by corresponding σ$^+$ intensity. (c) Degree of circularly polarzied PL $P$ as a function of pressure for (S-MBA)$_2$PbI$_4$ and (R-MBA)$_2$PbI$_4$.

In order to understand the underlying mechanism of the pressure-related increase in CPL degree, we investigated the structural evolution of (R-MBA)$_2$PbI$_4$ under pressure by synchrotron radiation XRD method. The angle dispersive x-ray diffraction (ADXRD) patterns under pressure up to 20.08 GPa are shown in Fig.4a. The position of diffraction peaks keeps increasing when the pressure is gradually increased, indicating that the lattice has been compressed. At the same time, the diffraction peaks broaden with increasing pressure, and thus some peaks become indistinguishable, indicating that the crystal lattice structure is getting disordered under high pressures. However, no other significant peaks or peak splitting were observed, suggesting that no structural phase transitions occurred during compression. When pressure increased to 15 GPa, the intensity of diffraction peaks decreased significantly and a broad background appeared, indicating that the crystal started to become amorphous.

The pressure dependence of the volume variation was also studied by using Birch-Murnaghan equation:

$$P(V) = \frac{3}{2}B_0\left[\left(\frac{V_0}{V}\right)^{\frac{7}{3}} - \left(\frac{V_0}{V}\right)^{\frac{5}{3}}\right]\left\{1 + \frac{3}{4}(B_0' - 4)\left[\left(\frac{V_0}{V}\right)^{\frac{2}{3}} - 1\right]\right\}$$

Where $B_0$ is bulk modulus and $B_0'$ is the derivative of $B_0$ with respect to pressure. The $B_0$ of the fitted curve is 18.6 GPa, larger than that of typical 3D perovskite like MAPbI$_3$ (13.6 GPa) and FAPbI$_3$ (11.0 GPa). Such a big bulk modulus contradicts the intuitive judgment that 2D perovskites are easier to get compressed, which could be derived from the strong anisotropic of the unit cell under high pressure. Similar phenomena have been reported and discussed in other achiral 2D perovskite. As a result, the strong anisotropic of the unit cell leads to ultrabroad PL and bandgap tunability.

Rietveld refinement of ARXRD patterns showed that the unit cell of (R-MBA)$_2$PbI$_4$ maintains the P2$_1$2$_1$2$_1$ space group when pressure is increased to 20.08 GPa, indicating that no structural phase transitions occurred during compression. The pressure-dependent unit cell parameters are shown in Fig.4b. We can find that the lattice parameters do not change much between 10.05 and 13.36 GPa, indicating an isostructural phase transition, which may lead to PL quenching at high pressure. It can be found that there is an obvious anisotropy in the lattice parameters when the pressure is applied. The variation along the direction perpendicular to the [PbI$_6$]$^{4-}$ octahedral layers ($\Delta c$) is three times greater than that in the in-plane directions ($\Delta a$, $\Delta b$). This anisotropy suggests that interlayer chiral organic molecules are more sensitive to pressure than the octahedral layers. The chirality of (S-MBA)$_2$PbI$_4$ and (R-MBA)$_2$PbI$_4$ perovskites is thought to result from chiral organic interlayers. From ADXRD results, the distance between chiral organic interlayer and [PbI$_6$]$^{4-}$ octahedral layer which corresponding to the PL emission decreases with the increase of pressure. Increased interlayer coupling and energy transfer between organic and Inorganic framework layers will lead to an enhanced overall chirality.

In addition to the interlayer coupling effects, the change of organic molecules under pressure will also leads to the overall chirality enhancement. However, it is difficult to accurately obtain the behavior of molecules in the organic layer from ADXRD results due to the complex structure of organic-inorganic hybrid perovskites. Raman spectroscopy was used to illustrate the change of chiral interlayers under high pressure, as shown in SI figure. Vibrational modes of the inorganic [PbI$_6$]$^{4-}$ octahedra are mainly active in the low frequency range (<200 cm$^{-1}$) due to the largely reduced mass, while the modes of organic cation appear at higher frequency (300–3200 cm$^{-1}$). At ambient pressure, three typical modes of organic cation are marked by comparing Raman spectrum of R-MBA, which located at 747, 1004 and 1035 cm$^{-1}$, respectively. These modes are continuously blueshift with the increasing pressure, which can be ascribed to the lattice contraction. It is worth noting that vibrational modes broaden with pressure and split at about 3 and 7 GPa, suggesting that the chiral organic layer undergoes structural distortion under compression. This structural distortion will enhance non-centrosymmetric properties of chiral organic cations, leading to an increased chirality, which will also induce the enhanced circularly polarized photoluminescence at high pressure.

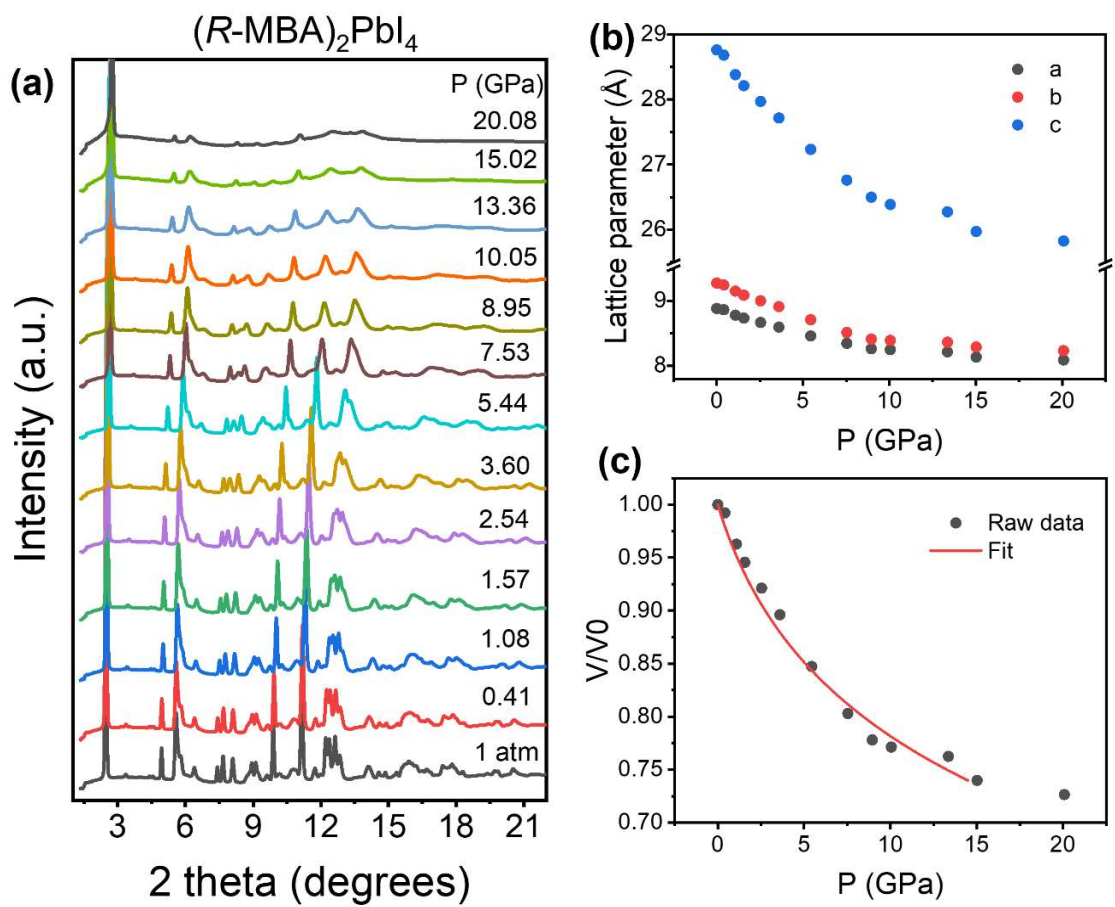

Figure 4. (a) Evolution of ADXRD pattern with pressure for (R-MBA)$_2$PbI$_4$. (b) Pressure dependence of unit cell parameters: a, b, and c. (c) Pressure dependence of unit cell volume. The volume change is represented in the form of V/V$_0$, in which V$_0$ is the initial volume at ambient pressure.

## 3. Conclusion

In summary, we studied the pressure effects on the chirality of chiral 2D perovskites (S- and R-MBA)$_2$PbI$_4$. The (S- and R-MBA)$_2$PbI$_4$ shown good optical tunability under pressure. With the increasing pressure, the PL intensity increased by about 10 times at 2 GPa, then gradually decreased until it could not be resolved beyond 9.2 GPa due to the structural amorphization. The PL emissions linearly redshifted from 517 nm at ambient pressure to 627 nm at 9.2 GPa. And the bandgap was decreased with pressure from 2.387 eV at 1 atm to 1.926 eV at 10.40 GPa, with a slope of about 44 meV/GPa. By using polarization-resolved photoluminescence measurements, the pressure-dependent CPL of (S- and R-MBA)$_2$PbI$_4$ perovskites were studied. The degree of CPL *P* was increased from nearly zero at ambient pressure to as high as 10% at 8.5 GPa. The significant increase in *P* value suggests that pressure can be used to tune the degree of the circularly polarized photoluminescence. ADXRD and Raman results indicated that the structural distortion and Increased interlayer coupling are responsible for the enhanced chirality under high pressure. Our findings provide a new approach to control the CPL of materials, showing potential applications in next generation of CPL devices.